\documentclass[aps,prl,floatfix,reprint,showpacs,superscriptaddress]{revtex4-1}
\usepackage[utf8x]{inputenc}
\usepackage{graphicx,epsfig,multirow,color}

\begin{document}

\title{Influence of steps on the tilting and adsorption dynamics of ordered pentacene films on vicinal Ag(111) surfaces}

\author{E. Mete}\email{emete@balikesir.edu.tr}
\affiliation{Department of Physics, Bal{\i}kesir University, Bal{\i}kesir 10145,
Turkey}
\author{\.{I}. Demiro\u{g}lu}
\affiliation{Department of Chemistry, Middle East Technical University, Ankara
06800, Turkey}
\author{E. Albayrak}
\affiliation{Department of Physics, Sakarya University, Sakarya, Turkey}
\author{G. Bracco}
\affiliation{CNR-IMEM and Department of Physics, University of Genoa, via
Dodecaneso 33, 16146 Genoa, Italy}
\author{\c{S}. Ellialt{\i}o\u{g}lu}
\affiliation{Department of Physics, Middle East Technical University, Ankara
06800, Turkey}
\author{M. F. Dan{\i}\c{s}man}\email{danisman@metu.edu.tr}
\affiliation{Department of Chemistry, Middle East Technical University, Ankara
06800, Turkey}

\date{\today}

\begin{abstract}
Here we present a structural study of pentacene (Pn) thin films on vicinal Ag(111)
surfaces by He atom diffraction measurements and density functional theory
(DFT) calculations supplemented with van der Waals (vdW) interactions. Our
He atom diffraction results suggest initial adsorption at the step edges 
evidenced by initial slow specular reflection intensity decay rate as a function 
of Pn deposition time. In parallel with the experimental findings, our 
DFT+vdW calculations predict the step edges as the most stable adsorption 
site on the surface. An isolated molecule adsorbs as tilted on the step edge with a 
binding energy of 1.4 eV. In addition, a complete monolayer (ML) with pentacenes
flat on the terraces and tilted only at the step edges is found to be more stable 
than one with all lying flat or tilted molecules, which in turn influences
multilayers. Hence our results suggest that step edges can trap Pn
molecules and act as nucleation sites for the growth of ordered thin films with
a crystal structure similar to that of bulk Pn.
\end{abstract}


\maketitle

Pentacene thin films are still being studied heavily due to their 
potential electronic device applications and being a model system for organic 
semiconductor film studies.\cite{anthony,facchetti,troisi} On many different
metal surfaces, Pn is reported to form lying down films with similar unit 
cell structures.\cite{kafer1,baldacchini,france1,guaino,kang1,kang2,
france2,kafer2,casalis,danisman1,pedio} In case of the second or higher layers 
however, even on the same metal surface, different growth mechanisms and/or 
dynamics have been reported.\cite{kafer1,kang1,kang2,beernink,eremtchenko,
dougherty} On Ag(111), while some reports suggest a bilayer film formation, 
where an ordered second layer with the symmetry of the Ag(111) 
surface forms on top of a disordered (2D gas phase) first layer at room 
temperature, some others suggest formation of bulk like Pn structures 
immediately after the first ML).\cite{kafer1,eremtchenko,dougherty} 
In addition, in our earlier works we had shown that an ordered Pn
multilayer could only be formed on a relatively high step density Ag(111) 
surface and the multilayer order can be improved by increasing the kinetic
energy (KE) of the molecules, by seeded supersonic molecular beam (SSMB) 
deposition.\cite{danisman1} This observation was attributed to step 
edges acting as nucleation centers for tilted molecules in the multilayers 
resulting in a step flow growth mechanism. Such growth dynamics was also
observed for different planar organic molecules on vicinal copper and gold
surfaces by different groups.\cite{gavioli,kamna,canas} In addition, high 
energy of the molecules was suggested to result in a local annealing effect 
which in turn helps formation of ordered multilayers at low substrate 
temperatures. These experimental results were in line with our recent 
theoretical work on Pn on flat Ag(111) surface where a multilayer 
structure with tilted Pn molecules was found to be more stable than one 
with flat lying molecules.\cite{mete} However during the above mentioned 
experimental studies  surface step density was changed arbitrarily by using a
miscut Ag(111) surface and the molecular KEs were not measured
by a time of flight (TOF) method but only estimated by using a theoretical
formula.\cite{danisman1} The necessity to test the stable configuration and
the contradictions in the literature have stimulated the present experimental
work on Pn film growth on Ag(11,12,12) surface complemented by
density functional theory (DFT) calculations to explore the effect of surface
step density and molecular energy on the Pn film growth with the aim
to improve our understanding of such systems. 

Pn films were grown by using SSMB deposition and characterized with a 
low energy atom diffraction (LEAD) apparatus both of which were detailed
before.\cite{casalis,danisman2} KE of the molecules was controlled by changing 
the nozzle temperature of the SSMB source and the carrier gas (CG). A commercial 
quartz crystal microbalance (QCM) and a quadruple mass spectrometer were 
implemented on this apparatus for allowing molecular flux and TOF measurements. 
In addition, a gold covered electrode of a custom design QCM can be used as the 
active substrate to simultaneously measure He specular intensity decay and 
record resonance frequency shift of the substrate to estimate film coverage 
during deposition.\cite{danisman3} Pentacene KEs, determined by TOF measurements, 
for standard SSMB source conditions were 2.2 eV for He CG and 0.2 eV for Kr CG. 
For LEAD measurements a monoenergetic He beam, with energy of 14 meV (incident 
wave vector $k_i$=5.13 {\AA}$^{\textrm{-}1}$) and velocity dispersion of 2\%, was scattered 
for the sample surface and the angular distribution of the scattering intensity was 
recorded. For reciprocal space mapping the diffraction spectra, recorded at 
80 K surface temperature along different crystallographic directions, were converted to 
momentum by using the equation $\Delta K_\parallel=k_i(\sin\theta_f- \sin\theta_i)$  
where $\Delta K_\parallel$ is the parallel momentum transfer, and $\theta_i$ and 
$\theta_f$ are the incident angle and the detector angular position, respectively. 
The reference ``flat'' Ag(111) and the Ag(11,12,12) substrates were purchased from 
Mateck GmbH with an orientation accuracy of $<$0.1$^\circ$ and had specular 
reflection intensity values of 60 \% and 45 \% of the incident beam respectively. 
Ag(11,12,12) has a miscut of 2.31$^\circ$ with respect to Ag(111) surface normal, 
with (111) terraces of 24 rows (58 {\AA}) wide with step edges parallel to [01\={1}] 
direction and perpendicular to [\={2}11] direction.

\begin{figure}[h!]
\epsfig{file=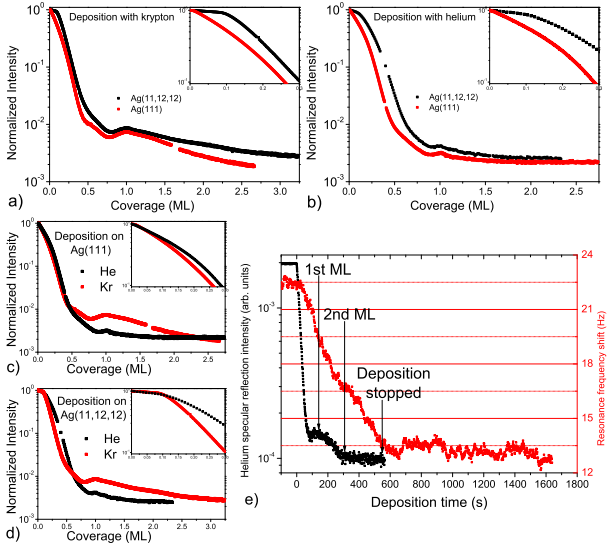,width=8cm,clip=true}
\caption{Specular reflection intensity decay curves as function of Pn
coverage.(a-d) Insets show the initial decay region. ($k_i$=5.13 {\AA}$^{\textrm{-}1}$,
incidence angle is 56$^\circ$ for Ag(111) and 59$^\circ$ for Ag(11,12,12), deposition 
temperature 200 K, on Ag(11,12,12) molecular beam direction was parallel to step edges) 
In (e) specular reflection intensity decay curve and resonance frequency shift from the 
gold electrode surface of a quartz crystal during Pn film deposition is shown.
\label{fig1}}
\end{figure}

\begin{figure}[ht]
\epsfig{file=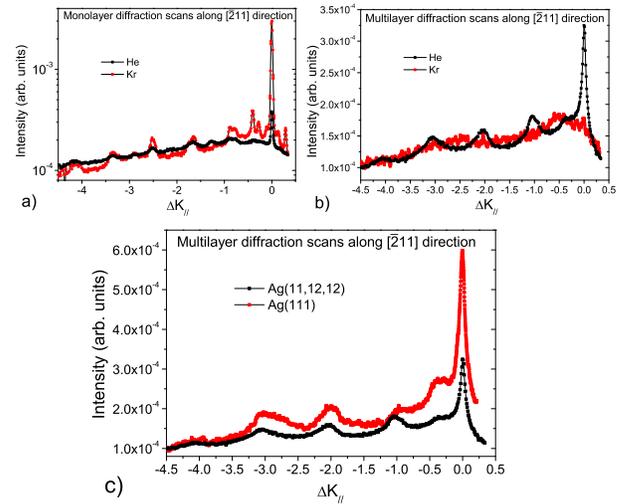,width=8cm,clip=true}
\caption{Helium diffraction scans of mono (a) and multilayers (b) of Pn
on Ag(11,12,12) with different KEs. Comparison of multilayer
diffraction scans on Ag(11,12,12) and Ag(111) grown with high KE are
shown in (c). ($k_i$=5.13 {\AA}$^{\textrm{-}1}$, incidence angle is 56$^\circ$ for
Ag(111) and 59$^\circ$ for Ag(11,12,12))\label{fig2}}
\end{figure}

In Fig.\ref{fig1} we show He specular reflection intensity decay curves as
a function of film coverage, at 200 K substrate temperature, for different
surfaces and deposition energies. The coverage values were deduced from the
specular reflection recovery peaks, with the assumption that these peaks 
correspond to ML completion. In fact, a similar recovery peak was
measured on Au(111) electrode surface of a quartz crystal at a resonance
frequency shift (2.9 Hz) corresponding to a deposited ML of
Pn as shown in Fig.\ref{fig1}e. When the initial part of these curves 
are examined it can be seen that, regardless of the type of the
CG, on Ag(11,12,12) there is a slow decay up to about 0.12 ML that is due 
to initial Pn adsorption to step edges. In fact, the He reflectivity
is very sensitive to molecular adsorbed species\cite{comsa} and this slow decay
can be attributed to the overlapping of Pn cross sections ($\sigma$) with that of
silver step edges resulting in a decrease of the effective cross section 
($\sigma_{\rm eff}$) of Pn molecules. However on Ag(111) due to lack of steps, 
Pn molecules mainly adsorb on the terraces and present a larger $\sigma_{\rm eff}$
to incoming He atoms which results in a faster decay of specular intensity. 
On the 24 rows wide terraces of Ag(11,12,12), 8 Pn rows with a 6$\times$3 unit 
cell are expected to adsorb at full coverage\cite{danisman1,mete} and a single 
Pn row is necessary to saturate the steps, corresponding to a coverage 1/8 
(0.125) ML. For higher coverage, Pn starts to fill terraces and the decay is 
faster explaining the observed specular intensity trend. In addition, when 
deposition is performed at 80 K substrate temperature, the slow decay region is 
completely missing due to reduced diffusion length of Pn molecules which prevent 
them to reach to the step edges. Regardless of the surface step density, the 
reflectivity decay curves for coverages of $>$0.3 ML show that deposition with
Kr yields a smoother ML as indicated by the higher specular reflection
intensity level. This can be due to disordering of the Ag step edges and terraces 
caused by the high energy of the Pn molecules deposited with He CG. In fact for 
Ag(111) surface, energies needed to detach an atom from a step edge (0.49-0.76 
eV) or to diffuse an atom along the step (0.29-0.34 eV)\cite{nandipati,chvoj}
are all much lower than Pn KEs deposited with He. Whereas in case of Kr CG, Pn 
molecules do not have enough energy to activate these roughening processes. 
In fact Pn $\sigma_{\rm eff}$s on Ag(11,12,12) (Fig~\ref{fig1}d), calculated by 
maximal attraction model proposed by Comsa\cite{comsa}, increases from 
$\sigma_{\rm Kr}=102 \pm 3$ {\AA}$^2$ to $\sigma_{\rm He}= 156 \pm 4$ {\AA}$^2$ 
for deposition with Kr and He respectively. When deposition was performed with 
a He beam perpendicular to step edges, $\sigma_{\rm He}$ further increases to 
328 $\pm$ 2 {\AA}$^2$, which indicates that step edge roughening is more effective 
in case of perpendicular deposition. These values can be compared with the Pn vdW 
cross section, $\sigma_{\rm vdW}$=124 {\AA}$^2$\cite{bondi}--119 {\AA}$^2$\cite{klein}. 
After the completion of the first ML, specularity decreases much faster in the case 
of Kr CG on both surfaces. This suggests that though the ML is more ordered/smooth 
with Kr, an ordered second layer does not grow on it. 

Diffraction scans, obtained from the MLs and the multilayers grown with
different KEs and on different substrates, shown in Fig.~\ref{fig2}
confirm the above mentioned mechanism:  On Ag(11,12,12) while the ML
diffraction peaks and specularity are more intense with Kr CG, no
multilayer diffraction peaks could be observed. A peak width analysis indicates
that average ML domain size with He beam is about 50 {\AA} whereas
with Kr it is about 70 {\AA} which means that with both CGs 
the domains extend as wide as the terrace width of the Ag(11,12,12) surface.
With He, however, multilayer diffraction peaks which are consistent with 
previously reported\cite{casalis,danisman1} Pn multilayer structure could 
be observed. Finally when the multilayer structure on Ag(111) and Ag(11,12,12) are
compared it can be seen that, though less intense, the diffraction peaks are
slightly narrower on Ag(11,12,12), corresponding to domain size of 17 {\AA} on
Ag(111) and 24 {\AA} on Ag(11,12,12). The reason for low specular reflection and
diffraction peak intensities on the vicinal surface may be the higher level of
diffuse scattering due to step edges, as discussed above.

\begin{table}[b!]%
\caption{DFT+vdW results of Pn on vicinal Ag(111) surface for different coverage
models. $E_{\rm t}$ is the relative supercell total energy.
Pn--step edge bond length is denoted by $l$. Labels denote the layer numbers for 
interlayer separations, $d_1$, $d_{1\textrm{-}2}$ and for tilting angles $\alpha_1$, 
$\alpha_2$, respectively.
\label{table1}}
\begin{ruledtabular}
\begin{tabular}{llccccccc}
 & Model & E$_{\rm t}${\footnotesize(eV)} & $l${\footnotesize({\AA})}
&$d_1${\footnotesize({\AA})} & $d_{1\textrm{-}2}${\footnotesize({\AA})} &
$\alpha_1${\footnotesize($^\circ$)} & $\alpha_2${\footnotesize($^\circ$)}
\\[1mm]\hline
\multirow{2}{*}{Single} & flat@bridge & ~0.00 & -- & 2.92 & -- & ~0.0 & --  \\
& C@step & -0.47 & 2.55 & -- & -- & 25.4 & -- \\[1mm] \hline
\multirow{2}{*}{1ML} & flat & ~0.00 & 2.63 & 2.94 & -- & ~0.0 & --  \\
& flat2\footnote{molecules flat on the terrace, tilted at the step edge.} &
-0.51 & 2.49 & 2.94 & -- & ~0.0 & -- \\[1mm] \hline
\multirow{2}{*}{2ML} & flat-tilt & ~0.00 & 2.54 & 2.77 & 2.56 & ~0.0 & ~17.0 \\
& flat2$^{\rm a}$-tilt &-0.49 & 2.50 & 2.77 & 2.56 & ~0.0 & -13.1 \\
\end{tabular}
\end{ruledtabular}
\end{table}

\begin{figure}[t1]
\epsfig{file=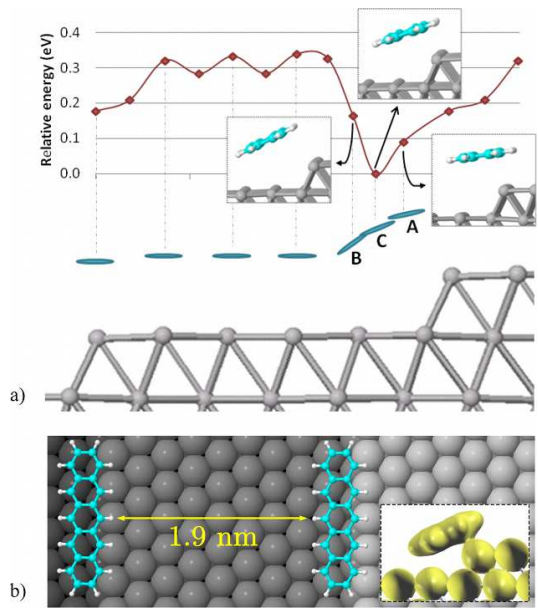,width=7cm}
\caption{(a) Potential energy profile of single Pn on Ag(233) vicinal
surface along [\={2}11] direction (A,B,C configurations shown in
the corresponding insets), (b) the minimum energy configuration of an isolated Pn 
at the step edge on Ag(455) surface (the charge
density is presented in the inset).\label{fig3}}
\end{figure}

\begin{figure}[t!]
\epsfig{file=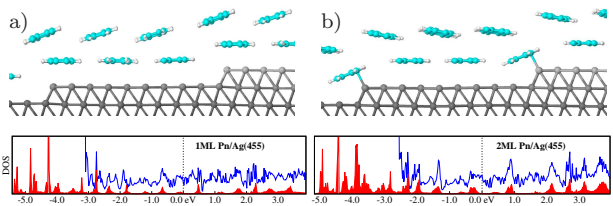}
\caption{Relaxed geomtries of 2ML Pn on Ag(455) for (a) flat-tilt and (b) 
flat2-tilt cases. Densities of states (DOS) are presented for the flat2 and 
flat2-tilt structures where the shaded regions are Pn contributions.   
\label{fig4}}
\end{figure}

To gain further insights on the system, we performed DFT calculations. Pure 
DFT methods underestimate the adsorption energy of Pn molecules on Ag(111). 
Among these, GGA-PW92 functional\cite{pw92} gives still low but slightly 
better binding energies and bond lengths for isolated and
ML cases on Ag(111) surface.\cite{mete} DFT calculations were
performed using VASP\cite{vasp} with projector augmented waves (PAW)
method\cite{paw1,paw2}. We employed Grimme's semiempirical forcefield
approach\cite{grimme} to include long range vdW corrections to better describe
the interaction of Pn molecules with Ag substrate. We used a 4 layer slab to
model the vicinal Ag(111) surface with two different terrace sizes. The width of
the terraces on Ag(233) and Ag(455) supercells measure 10.0 {\AA} and 20.0 {\AA} 
spanning six and ten surface Ag atoms along [\={2}11], respectively. Ag(233) 
supercell has been used to study the potential energy profile of an isolated Pn 
as schematically shown in Fig.\ref{fig3}a.  Ag(455) slab for one and two ML
coverages allows to place three Pn molecules one being at the step edge as depicted 
in Fig.\ref{fig4}. We have relaxed all probable configurations based on
minimization of the forces requiring them to be smaller than 0.1
eV$\,${\AA}$^{\textrm{-}1}$ on every atom. In order to elucidate the effect of van
der Waals contribution on Pn-metal interaction that characterizes the minimum
energy structures, we carried out these calculations both with and without
dispersive energy corrections.

Pn molecules give noticeably stronger binding at the step edges relative to 
that on the flat terraces as seen in Fig.\ref{fig3}a which shows the diffusion
barrier profile along [\={2}11] obtained with DFT. Pn at the step edge bends 
along its long axis such that the middle part of the molecule gets slightly 
more closer to the surface from where we measure Pn--step edge bond length 
(column $l$ in Table~\ref{table1}). The curvature is similar to the one observed 
for Pn on Cu(111) surface.~\cite{smerdon} DFT+vdW reproduces this curvature that 
indicates a strong adsorption. Binding energies at the step edge 
(case C in Fig.\ref{fig3}a) and at the bridge site (molecular center on 
a Ag-Ag bond) on the terrace are 1.40 and 0.92 eV, respectively. 
These were much lower with pure DFT. For instance, DFT predicts the binding
energy at the step edge to be 0.62 eV with no bending.

For the first ML, we have considered tilted, flat and flat2 models as initial
structures for geometry optimization. In tilted case all molecules are tilted
around their major axes. Flat model is as the first layer of Fig.\ref{fig4}a 
where all Pn molecules lie parallel to the Ag surface plane. Flat2 case differs
from the flat one in only that the Pn at the step edge is tilted as in
the most stable case C of an isolated molecule (see the first layer of 
Fig.\ref{fig4}b). Pure DFT predicts tilted formation to be the minimum energy 
configuration while DFT+vdW relaxes tilted case to flat2 one. Moreover, the 
measured $\sigma_{\rm eff}$s provide a further support to flat2 model because 
Pn covers completely a step edge and the tilting reduce the $\sigma$ to 
$\sigma_{\rm vdW}$ $\cos(25.4)=112$ {\AA} which is at most 10\% greater than 
$\sigma_{\rm Kr}$. In the flat model, the $\sigma_{\rm vdW}$ is increased by 
the step edge cross section, so, apparently the $\sigma_{\rm He}$ supports the 
flat case. But the experimental evidence that He deposition causes an increased 
disorder and further calculations that exclude the presence of other 
configurations with minimal energy similar to flat2, leaves flat2 as the only 
possibility.

Then we considered a second ML which can be tilted or flat as an initial 
configuration.  For 2ML Pn on Ag(455), pure DFT predicts tilted--tilted 
model to be the lowest energy structure. When the geometry optimization 
repeated with dispersive corrections, tilted molecules on the terrace become 
flat leading to flat2-tilt configuration (Fig.\ref{fig4}b). Therefore, second 
ML is tilted while first layer lies flat similar to that of the 1ML case. This 
is in good agreement with the proposed model for bilayer Pn on Cu(111) from 
STM/STS measurements.\cite{smerdon}  This indicates the role of the step edge 
in bulk--like formation of Pn molecules beyond the second ML. As a trend, 
energy differences between flat and flat2 cases for both 1ML and 2ML coverages 
follow from that of single molecule adsorption cases, \textit{i.e.} flat@terrace 
and C@step as presented in Table~\ref{table1}. Moreover, a comparison of DOS 
structures for Ag(455) in Fig.\ref{fig4} with those obtained for Ag(111) without 
steps\cite{mete} shows that the difference between the HOMO level of Pn 
and the Fermi energy of the surface decreases. This suggests an enhanced charge 
carrier injection at Pn--metal interface in favor of the vicinal surface.   

In summary, by using He diffraction and QCM techniques 
simultaneously and performing DFT+vdW calculations we have, unambiguously, 
shown that step edges on Ag(111) surfaces act as nucleation centers for
formation of bulk--like Pn thin films on MLs composed of flat
lying molecules. This mechanism can be considered to be a rather general one
since our results are in line with those reported on different metal surfaces
such as Au and Cu. Our results highlight the importance of a) vdW forces for
systems with relatively weak interactions like Pn/metal interfaces and b)
the effects of metal substrate surface morphology on the crystal and electronic
structure of organic semiconductor films which have very important implications
on the performance of electronic devices that employ these metal/organic
interfaces.

This work was partially supported by T\"{U}B\.{I}TAK Grant Nos. 107T408 and 209T084.

\end{document}